\documentstyle {article}

\begin{document}

{\bf Vortex lattice melting theories as example of science fiction.}

\

A.V.Nikulov,

Institute of Microelectronics Technology and High Purity Materials, Russian
Academy of Sciences, 142432 Chernogolovka, Moscow District, Russia.

\

\begin{abstract}It is shown that the popular conception of the vortex
lattice melting have appeared in consequence of a incorrect notion about
the Abrikosov state and incorrect definition of the phase coherence. The
famous Abrikosov solution gives qualitatively incorrect result. The
transition into the Abrikosov state must be first order in ideal (without
disorder) superconductor. Such sharp transition is observed in bulk
superconductors with weak disorder below $H_{c2}$. No experimental evidence
of the vortex lattice melting exists now. The absence of the sharp
transition in thin films with weak disorder is interpreted as absence of
long-rang phase coherence down to very low magnetic field. The observed
smooth phase coherence appearance in superconductors with strong disorder
is explained by increasing of the effective dimensionality. It is proposed
to return to the Mendelssohn model for the explanation of the resistive
properties of superconductors with strong disorder. It is conjectured that
the Abrikosov state is not the vortex lattice with the crystalline
long-rang order.
\end{abstract}

\

\section{Introduction}

	I think nobody will contest that some theoretical works is sciences
fiction only. It can be easy proved. Some different points of view exist
about many problems. But a one only or none of them correspond
to a reality. Consequently, other points of view are science fiction only.
I understand this situation is inevitable. Theorist fantasy is much wider
than a reality. It is very difficult sometimes to tell a fiction from
the reality. But we must strive for this. And there is very important a
frank discussion.

	In the present work I try to convince of readers that the conception of
the vortex lattice melting in the mixed state of type II superconductors,
which was very popular the last ten years and continues to be popular now,
is sciences fiction only.

\section{Why has been appeared and become popular the conception of vortex
lattice melting?}

	The conception of vortex lattice melting has been appeared and became
popular because the point of view prevailed that the Abrikosov state is the
flux line (or vortex) lattice (FLL) \cite{brandt95} like to atom lattice, or
lattice of long molecules \cite{nelson95}. And now, almost nobody doubt
that it is so. This opinion was caused by direct observation of the vortex
lattice \cite{trauble}.

	The Abrikosov vortex was considered as a magnetic flux. The transport
properties were connected with the motion of the magnetic flux structures
(or FLL) \cite{huebener}. According to all textbook (see for example
\cite{huebener}), the steady-state motion of the magnetic flux structure
causes the time-averaged macroscopic electric field following Faraday's
law. Motion of the magnetic flux structure can be induced by the Lorentz
force. The resistivity in the Abrikosov state is called "flux flow
resistivity" \cite{huebener}.

	The Lorentz force is compensated by the damping force and the pinning
force. The pinning force is caused because the vortexes are pinnid by
superconductor disorder. The vortex pinning is very important effect.
Application of type II superconductors in high magnetic
fields is possible owing to the pinning effect only. Therefore it was
strong disappointment \cite{bishop} when soon after the high-Tc
superconductor discovery one has detected that the pinning effect is absent
in a wide region below the second critical field, $H_{c2}$. This was
interpreted as consequence of the vortex lattice melting because the
Abrikosov state was considered as the flux line (or vortex) lattice.

	According to popular point of view \cite{bishop}, the process of the
vortex pinning is very different for vortex liquids or solids. In the case
of a vortex solid, a few pinning centers can hold the entire lattice
because it is stiff. But it is impossible to hold in place a vortex liquid
with help a few pinning centers.

\section{The notions, on which the conception of the vortex lattice melting
is founded, are no quite right.}

\subsection{The Abrikosov vortex is not a flux line but is a singularity in
the mixed state with the phase coherence.}

	The Abrikosov vortices appear because a magnetic flux cannot be within a
superconducting region with long-rang phase coherence and without
singularities. The later is followed from the relation for the
superconducting current \cite{huebener}

$$\frac{\Phi_{0}}{2\pi}\int_{l}dR \frac{d\phi}{dR} =
\int_{l}dR\lambda_{L}^{2}j_{s} + \Phi \eqno{(1)}$$

where l is a closed path of integration; $\lambda_{L} =
(mc/e^{2}n_{s})^{0.5}$ is the London penetration depth; $j_{s}$ is the
superconducting current density; $\Phi$ is the magnetic flux contained
within the closed path of integration l. If the singularity is absent
$\int_{l}dR d\phi /dR = 0$. In this case the relation (1) is the equation
postulated by F. and H.London \cite{london35} for the explanation of the
Meissner effect \cite{meissner} (see \cite{huebener}).

	According to (1) a magnetic field can penetrate within a superconductor
only if: 1) superconductivity is destroyed, or 2) singularities appears, or
3) the long-rang phase coherence is absent. The first is observed in type I
superconductors. The second is the case of the Abrikosov state. The
existence of the vortices is evidence of the existence of the phase
coherence. Consequently, the Abrikosov state is the mixed state with
long-rang phase coherence. $(\int_{l}dR d\phi/dR)/2\pi = n$ is a number of
the vortices contained within l. If the l radius $\gg \lambda_{L}$ then $n
= \Phi/\Phi_{0}$. The third case is a mixed state without the long-rang
phase coherence.

	\subsection{The resistivity in the Abrikosov state is caused by the vortex
motion but no the motion of the magnetic flux.}

	The time-averaged macroscopic electric field appears in the Abrikosov
state in accordance no with the Faraday's law but with the Josephson's
relation

$$V = \frac{\hbar}{2e}\frac{d\phi}{dt} \eqno{(2)}$$
The vortex flow causes a change of the phase difference $\phi$ in time and
as consequence of (2) a voltage. The Faraday's law and the Josephson
relation give the same result. Therefore it has became possible that the
resistivity in the Abrikosov state is considered as flux flow resistivity
in all textbook and majority of papers, although it is obvious that the
magnetic flux does not flow in a superconductor. This no right notion is
one of the causes of the wide popularity of the conception of the vortex
lattice melting. I will use the more right denomination "vortex flow
resistivity" instead of "flux flow resistivity".

\subsection{The vortex pinning is a consequence of the long-rang phase
coherence.}

	The vortex pinning is a consequence of the long-rang phase coherence,
because the Abrikosov vortex can not exist without the phase coherence.
Consequently, the pinning disappearance can be interpreted as the phase
coherence disappearance. It is important to note this because most authors
interpret the pinning disappearance as the vortex lattice melting.

	The vortex pinning has an influence on resistive properties first of
all. The resistivity has different nature in the states with and without
the phase coherence. Consequently, resistive properties change first of all
at the long-rang phase coherence appearance transition.

\subsection{No evidence exists now that the Abrikosov state is the vortex
lattice with the crystalline long-rang order.}

	Many scientists think that the direct observation \cite{trauble} is
evidence that the Abrikosov state is the vortex lattice. But it is not
right. If we lay along a fishing net with help of stakes it will look as a
lattice. But from this direct observation we can not draw a conclusion
that the fishing net is a lattice which can melt.

	A real vortex lattice is a structure in a inhomogeneous space, because
disorders are in any real superconductor sample.  Larkin \cite{larkin70}
has shown that the crystalline long-rang order of the vortex lattice is
unstable against the introduction of random pinning.  Consequently we can
not contend on base of direct observation \cite{trauble} that the Abrikosov
state is the vortex lattice which can melt because it can be a structure
like the fishing net.

	Some theorists found the vortex lattice melting theories on the
Abrikosov solution \cite{abrikos} and posterior results \cite{fetter67}.
According to the mean field approximation the Abrikosov state is the
triangular vortex lattice with the crystalline long-rang order indeed
\cite{kleiner}. But according to \cite{maki71} the mean field
approximation can not be used for the description of the mixed state in the
thermodynamic limit.

	Maki and Takayama \cite{maki71} have shown that the fluctuation
correction $\Delta n_{s,fl}$ to the Abrikosov solution calculated in the
linear approximation depends on superconductor size L across magnetic field
direction: $\Delta n_{s,fl}$ is proportional to $\ln(L/\xi)$ in
three-dimensional superconductor and $\Delta n_{s,fl}$ is proportional to
$(L/\xi)^{2}$ in two-dimensional superconductor. This result seems very
queer for most scientists, because they think that it contradicts to
experimental results. Therefore almost nobody has believed in a reality of
this result, even authors. In order to "correct" this result Maki (with
Thompson) have made incorrect work \cite{maki89}.

	In spite of the opinion of most scientists I claim that the Maki-
Takayama result \cite{maki71} is right. It does not contradict to direct
observation of the Abrikosov state. But according to this result the
thermal fluctuation in the mixed state can not be considered as
perturbations in the thermodynamic limit. Therefore the Abrikosov solution
can not be used for a basing of the vortex lattice melting theories.

\section{Theories of the vortex lattice melting and theories of the vortex
liquid solidification.}

	Soon after the HTSC discovery the elastic theories of the vortex lattice
melting appeared \cite{nelson89}. The vortex lattice
melting is considered as consequence of increasing of thermal displacement
of the vortex position in these theories. In analogy to crystal lattice a
melting temperature $T_{m}$ was estimated from the Lindeman criterion
\cite{lindeman}.

	Dislocation-mediated melting was discussed in some works
\cite{fisher80}. Numerous other theories on a possible melting transition
were published (see review \cite{brandt95, blatter}). All these theories
are based on the assumption that the Abrikosov state is the vortex (or flux
line) lattice with crystalline long-rang order.

	The elastic theories of vortex lattice melting raised doubts soon after
they have appeared, because the description in these theories is
unsatisfactory in principle, since it starts from the state in which the
translation symmetry has been broken by hand \cite{zt91dec}. Therefore some
theorists consider no the vortex lattice melting but the solidification
transition of vortices. The solidification theories are a revision of the
Abrikosov solution \cite{abrikos}. They try to find and to describe the
transition into the vortex lattice state, taking into account the thermal
fluctuation.

	First attempt to find the solidification transition was made before the
HTSC discovery \cite{rugger76}. No transition was found in the approaches
\cite{rugger76,brezin90} based on perturbation theory. Thus, the transition
into the Abrikosov state is lost in perturbation fluctuation theory.  But
because it is well known from the direct observation that the Abrikosov
state exists the attempts to find the solidification transition were
continued in many works in different approaches. Most authors find the
solidification transition \cite{tesan91}. And few authors \cite{moore}
only ventured to state that the solidification transition is absent.

	I agree with M.A.Moore that the solidification transition is absent. But
this does not mean that the transition into the Abrikosov state is absent,
because the Abrikosov state is the mixed state with long rang phase
coherence. The solidification theories consider no the long-rang phase
coherence appearance but the destruction of the translation symmetry
because a wrong definition of the phase coherence is used in these
theories.

\section{Definition of the phase coherence.}

	The phase coherence is defined by the correlation function in the
solidification theory. According to this definition the long-rang phase
coherence can not exist without the crystalline long-rang order of the
vortex lattice. It is claimed in some works \cite{ikeda96} that the
phase coherence can remain short-ranged even in the vortex solid phase.

	The phase coherence definition used in the solidification theory is
logical contradictory because the existence of the vortexes is evidence of
the long-rang phase coherence. And this must follow from the right
definition. We can use the relation (1) for the definition of the phase
coherence: the phase coherence exists in some region if the relation (1)
is valid for any closed path in this region.

	According to the right definition the long-rang phase coherence must be
both in the vortex lattice and in the vortex liquid. Consequently a phase
coherence disappearance transition must be observed above the vortex
lattice melting.

\section{Because I think that the vortex lattice melting is science
fiction.}

	But the only transition is observed. This transition exists for certain
in bulk superconductors with weak disorder. It is observed first of all at
the investigation of the resistive dependencies in a perpendicular magnetic
field. First this transition was observed in conventional superconductors
\cite{nik81l} before the HTSC discovery. Later, this result was repeated
in many works \cite{welp} by investigations of $YBa_{2}Cu_{3}O_{7-x}$
single crystals with weak disorder. This transition is observed below
$H_{c2}$ at a magnetic field denoted as $H_{c4}$ in \cite{nik90}. A
difference of the $H_{c2} - H_{c4}$ values observed in conventional
superconductors \cite{nik81l} and in $YBa_{2}Cu_{3}O_{7-x}$ \cite{welp}
conforms to the scaling law \cite{nik96}.

	This transition was interpreted in our work \cite{nik81l} as a phase
transition from a "one-dimensional" state (the mixed state without the
long-rang phase coherence) into the Abrikosov state. But most authors
interpret this transition as the vortex lattice melting \cite{welp}. This
interpretation can not be right because no transition is observed above.
According to this interpretation the vortex liquid exists above $H_{c4}$.
But no experimental evidence of the vortex liquid exists now. Moreover, our
investigation of bulk conventional superconductors \cite{nik85} and thin
films \cite{nik95l,nik97} show that the vortex liquid does not exist.

	Some properties of the vortex liquid must differ from the one of the
mixed state without the phase coherence. For example, a nonlocal
resistivity must be observed in the vortex liquid \cite{nonlocal} and be
not observed in the mixed state without the phase coherence. Our
investigations of nonlocal conductivity in thin films of conventional
superconductor \cite{nik97} show that the magnetic field destroys the
phase coherence first of all. In a wide region below $H_{c2}$ the phase
coherence is absent. Therefore the opinion of most authors that the
observed transition is the vortex lattice melting and the state between
$H_{c4}$ and $H_{c2}$ is the vortex liquid can not be right.

\section{The famous Abrikosov solution gives qualitatively incorrect
result.}

	According to the Abrikosov solution \cite{abrikos} a second order phase
transition from the normal state into the superconducting state with the
long-rang phase coherence takes place in $H_{c2}$. It is right in the mean
field approximation. But it is no right in fluctuation theory.

	The long-rang phase coherence appearance may be considered as a
consequence of the increasing up to the infinity of the coherence length.
The correlation length becomes anisotropic in a high magnetic field
\cite{tinkha75}. The longitudinal (along a magnetic field) coherence length
calculated in the linear approximation $\xi_{l} = (\Phi_{0}/2\pi
(H-H_{c2})^{0.5}$ increases up to infinity at $H_{c2}$, whereas the
transversal coherence length $\xi_{t} = (2\Phi_{0}/\pi H)^{0.5}$ changes
little near $H_{c2}$. The longitudinal coherence must be renormalized near
$H_{c2}$, whereas the transversal coherence length changes little in the
critical region. This means that the correlation function of bulk
superconductor near $H_{c2}$ (in the lowest Landau level (LLL)
approximation region) is similar to the one of one-dimensional
superconductor \cite{lee72}. Consequently, if we define the phase coherence
in the mixed state by the correlation function we can conclude that the
long-rang coherence can not be in the mixed state of type II
superconductors.

	On other hand we know that the Abrikosov state is the mixed state with
long-rang phase coherence (with the length of phase coherence equal sample
size L). The mixed state with the length of the phase coherence across
magnetic field direction $(\Phi_{0}/H)^{0.5}$ exists also. I call this
state as the mixed state without the phase coherence (or "one-dimensional"
state). Two characteristic lengths only ($(\Phi_{0}/H)^{0.5}$ and sample
size L) are in a superconductor without disorder.  Consequently, a length
of the phase coherence in the ideal superconductor can change only by jump
from $(\Phi_{0}/H)^{0.5}$ to L. I.e the transition into the Abrikosov
state must be first order (sharp) phase transition in ideal case.

\section{The position of the transition into the Abrikosov state depends on
disorder amount in thin films.}

	The sharp transition is observed at $H_{c4}$ in bulk superconductors
with weak disorder \cite{nik81l,welp}. The results of work \cite{schiling}
show that it may be the first order phase transition indeed. But no sharp
transition is observed in thin films \cite{nik95l,kes97} and in bulk
superconductors with strong disorder \cite{fendrich}.

	The absence of any features of the resistive properties up to low
magnetic field in amorphous $Nb_{1-x}O_{x}$ films was interpreted in our
paper \cite{nik95l} as the absence of the transition into the Abrikosov
state. This interpretation was criticized by Theunessen and Kes
\cite{kes97}. They contend that we do not observe any feature because our
measuring current is extremely high in comparison to the critical current.
But sharp feature of the vortex flow resistivity must be observed also at
the transition into the Abrikosov state in superconductors with weak
disorder according to the fluctuation theory \cite{maki71}.

	Such sharp features are observed in all enough homogeneous bulk
superconductors \cite{nik81j, nik85}. And only in no enough homogeneous
samples the "classical" flux flow resistivity dependencies \cite{kim64} are
observed. The observed feature differ qualitatively from the mean-field
vortex flow resistivity dependence \cite{kopnin}.

	Smooth features of the vortex flow resistivity are observed in a middle
field in the a-NbGe films with an intermediate strength of disorder
\cite{kes97} (see the inset of Fig.7). This feature is a consequence of the
phase coherence appearance. But no features is observed down to very low
fields in the $Nb_{1-x}O_{x}$ films with extremely small pinning
\cite{nik95l}. The features of the vortex flow resistivity can be observed
at high measuring current. Therefore I can claim that the phase coherence
appearance in thin film depends on the amount of disorder. This claim is
confirmed by my theoretical results \cite{nik95b}.

\section{Transition into the Abrikosov state in superconductors with strong
disorder. The return to the Mendelson model.}

	The transition into the Abrikosov state in superconductors with strong
disorder \cite{kes97,fendrich} is smooth. The length of the phase
coherence does not change by jump but increases gradually with the magnetic
field (or the temperature) decreasing \cite{nik98}.

	To explain this difference from the ideal case I propose in
\cite{nik98} to return to Mendelssohn model \cite{mendelss}. The
Mendelssohn's sponge \cite{mendelss} can be considered as a limit case of
strong disorder. Real superconductors can be considered as intermediate
cases between the Mendelssohn's \cite{mendelss} and Abrikosov's
\cite{abrikos} models. The Mendelssohn sponge is a one-dimensional system.
Therefore a thin film with strong disorder can be considered as a system
like the one-dimensional superconductor. In one-dimensional superconductor
\cite{grunberg} the length of the phase coherence increasing smoothly with
temperature decreasing. In consequence of this the resistive transition is
smooth also.

\section{Conclusions}

	The Abrikosov state is the mixed state with long-rang phase coherence.
Consequently two long-rang orders exist in the Abrikosov state if it is the
vortex lattice with crystalline long-rang order. But the only transition is
observed on the way from the Abrikosov state into the normal state.
Consequently, or both orders disappear at this transition, or the Abrikosov
state is not the vortex lattice. I conjecture that the second corresponds to
the reality. Observed correlation between the vortex lattice and the
crystal lattice of superconductor \cite{huebener} confirms my opinion.

	The second critical field $H_{c2}$ is no critical point not only in
superconductors with weak disorder but also in superconductors with strong
disorder. The phase coherence appears below $H_{c2}$ in superconductors
with weak disorder \cite{nik81l,welp} and above $H_{c2}$ in superconductors
with strong disorder \cite{nik98}.

	The sharp transition into the Abrikosov state predicted by the
fluctuation theory in ideal case is observed in bulk superconductors with
weak disorder \cite{nik81l,welp} only. No sharp transition is observed in
thin films with weak disorder \cite{nik95l}. This difference can be
explained by difference of the fluctuation value in three- and
two-dimensional superconductors \cite{maki71}.

	The smooth phase coherence appearance in superconductors with strong
disorder can be explained qualitatively by the increasing of the effective
dimensionality of the fluctuation.

\section{Acknowledgment}

	I thank for financial support the International Association for the
Promotion of Co-operation with Scientists from the New Independent States
(Project INTAS-96-0452) and the National Scientific Council on
"Superconductivity" of SSTP "ADPCM" (Project 98013).

\end{document}